\documentclass[10pt,conference,compsoc]{IEEEtran}

\usepackage[nocompress]{cite}
\usepackage[pdftex]{graphicx}
\usepackage{amsmath}
\usepackage{array}
\usepackage{caption}
\usepackage[caption=false,font=footnotesize,labelfont=sf,textfont=sf]{subfig}
\usepackage{url}
\usepackage{color}
\usepackage{helvet}

\begin{document}

\title{Stop Clickbait: Detecting and Preventing Clickbaits in Online News Media}
\author
{
\IEEEauthorblockN{Abhijnan Chakraborty, Bhargavi Paranjape, Sourya Kakarla, Niloy Ganguly}
\IEEEauthorblockA{Department of Computer Science and Engineering\\
Indian Institute of Technology Kharagpur, India -- 721302}
}

\maketitle

\begin{abstract}
Most of the online news media outlets rely heavily on the revenues generated from the clicks made by their readers,
and due to the presence of numerous such outlets, they need to compete with each other for reader attention. 
To attract the readers to click on an article and subsequently visit the media site, the outlets
often come up with catchy headlines accompanying the article links,
which lure the readers to click on the link. 
Such headlines are known as \textit{Clickbaits}. While these baits
may trick the readers into clicking, in the long-run, clickbaits usually don't live up to the 
expectation of the readers, and leave them disappointed.

In this work, we attempt to automatically detect clickbaits and 
then build a browser extension which warns the readers of different media sites 
about the possibility of being baited by such headlines.
The extension also offers each reader an option to block clickbaits she doesn't want to see. 
Then, using such reader choices, the extension automatically blocks similar
clickbaits during her future visits. We run extensive offline and online experiments across multiple media sites
 and find that the proposed clickbait detection and the personalized blocking approaches perform very well 
 achieving $93\%$ accuracy in detecting and $89\%$ accuracy in blocking clickbaits.
\end{abstract}

\vspace{-0.2cm}
\section{Introduction}
\vspace{-0.3cm}
\noindent 
With the news consumption gradually moving online, the media landscape is undergoing a sea change. We can attribute this change to two primary dimensions. First, compared to the traditional offline media, where the readers' allegiances to a particular newspaper were almost static, online media offers the readers a gamut of options ranging from local, national or international media outlets to several niche blogs specializing on particular topics of interest. Second, most of the online media sites do not have any subscription charges and their revenue mostly come from the advertisements on their web pages.

Essentially, in the online world, every media outlet has to compete with many such outlets for reader attention and make their money from the \textit{clicks} made by the readers. Therefore, to attract the readers to visit the media site and 
click on an article, they
employ various techniques, 
such as coming up with catchy headlines accompanying the article links, which lure the readers to click on the links. Such headlines are known as \textit{Clickbaits}. 
According to the Oxford English Dictionary, clickbait is defined\footnote{\url{oxforddictionaries.com/us/definition/american_english/clickbait}} as ``(On the Internet) content whose main purpose is to attract attention and encourage visitors to click on a link to a particular web page."
Examples of such clickbaits include \textit{``This Rugby Fan's Super-Excited Reaction To Meeting Shane Williams Will Make You Grin Like A Fool", ``15 Things That Happen When Your Best Friend Is Obsessed With FIFA", ``Which Real Housewife Are You Based On Your Birth Month"} or the epic \textit{``They Said She Had Cancer. What Happens Next Will Blow Your Mind"}.

Clickbaits exploit the cognitive phenomenon known as \textit{Curiosity Gap}~\cite{loewenstein1994psychology}, 
where the headlines provide forward referencing cues to generate enough curiosity 
among the readers such that they become compelled to click on the link to fill the knowledge gap. 
While these baits may trick the readers into clicking, in the long-run, clickbaits usually don't live up to the 
expectation of the readers, and leave them disappointed. 
Cognitive studies (such as~\cite{clickbait_attention}) have argued that clickbait is an enabler of {\it attention distraction}. 
As the readers keep switching to
new articles after being baited by the headlines, the attention residue from these constant switches
 result in cognitive overload, deterring the readers from reading more informative and in-depth news stories.
There are also concerns regarding the role of journalistic gatekeeping in the changed media landscape with 
the prevalence of clickbaits~\cite{clickbait_bad}.

Even with all these hue and cry around the ill effects of clickbaits, 
there has been little attempt to devise a systematic approach for a comprehensive solution. 
In 2014, Facebook declared that they are going to remove clickbait stories from users' news feeds\footnote{newsroom.fb.com/news/2014/08/news-feed-fyi-click-baiting}, depending on the click-to-share ratio and the amount of time spent on these stories.
Yet, Facebook users still complain that they continue to receive clickbaits and there is a renewed effort to clamp down on clickbaits\footnote{thenextweb.com/facebook/2016/04/21/facebook-might-finally-kill-clickbait-new-algorithm-tweaks/}. 
In a recent work, Potthast et al.~\cite{potthast2016clickbait} attempted to detect clickbaity tweets in Twitter.
The problem with such standalone approaches is that clickbaits are prevalent not only on particular social media sites,
 but also on many other reputed websites across the web. 
 For example, the `Promoted Stories' section at the end of the articles in the websites of `The Guardian', 
 or `Washington Post' contain many clickbaits. 
Therefore, we need to have a comprehensive solution which can work across the web.

There have been some ad-hoc approaches like `Downworthy'~\cite{downworthy} which detects clickbait headlines using 
a fixed set of common clickbait phrases and then converts the headlines into something more garbage-ish, 
or `Clickbait Remover for Facebook'~\cite{clickbait_remover} 
which prevents the links to a fixed set of domains from appearing in the users' news feeds. 
The problem with having a fixed rule set is they are not scalable and may need constant tuning with the emergence of 
new clickbait phrases. 
Similarly, preventing links to a fixed set of domains will also block article links which are not clickbaits.

In this work, we take the first step towards building a comprehensive solution which can work across the web. 
We first build a classifier which automatically detects whether a headline is clickbait or not. Then we explore ways to block
certain clickbaits from appearing in different websites. A survey conducted on $12$ regular readers of news media sites
suggested that the headlines the readers would like to block vary greatly across the readers, 
and they are influenced by the particular reader's interests. 
Hence, instead of a generalized solution, we develop personalized classifiers for individual readers which can
predict whether the reader would like to block a particular clickbait given her earlier block and click history.

We finally build a browser extension, `Stop Clickbait', which warns the readers about the possibility of being baited by 
clickbait headlines in different media sites.
The extension also offers the readers an option to block certain types of clickbaits they would not like to see during future encounters. We run extensive offline and online experiments across multiple media sites and find
that the proposed clickbait detection and personalized blocking approaches perform very well 
achieving $93\%$ accuracy in detecting
and $89\%$ accuracy in blocking clickbaits. We believe that the widespread use of such extensions would deter the readers from 
getting lured by clickbait headlines, which in turn would disincentivize the media outlets from relying on clickbaits
as a tool for attracting visitors to their sites.

\vspace{-0.2cm}
\section{Dataset}
\vspace{-0.3cm}
\noindent We collected extensive data for both clickbait and non-clickbait categories.

\vspace{0.1cm}
\noindent\textbf{Non-clickbait:} We extracted the headlines from a corpus of $18,513$ Wikinews articles collected 
by NewsReader~\cite{newsreader2016kbs}. In Wikinews, articles are produced by a community of contributors and 
each news article needs to be verified by the community before publication. 
There are fixed style guides which specify the way some events need to reported and presented to the readers. 
For example, to write the headline of a story, there are a set of guidelines\footnote{\url{en.wikinews.org/wiki/Wikinews:Style_guide#Headlines}} the author needs to follow. Due to these rigorous checks employed by Wikinews, we have considered the headlines of these articles as gold standard for non-clickbaits.
 
\vspace{0.1cm}
\noindent\textbf{Clickbait:} For clickbaits, we manually identified the following domains which publish many clickbait articles: `BuzzFeed', `Upworthy', `ViralNova', `Scoopwhoop', and `ViralStories'. We crawled $8,069$ web articles from these domains during the month of September, 2015. To avoid false negatives (i.e. the articles in these domains which are not clickbaits), we recruited six volunteers and asked them to label the headlines of these articles as either 
clickbait or non-clickbait. We divided the articles among the volunteers such that 
each article is labeled by at least three volunteers. 
We obtained a `substantial' inter-annotator agreement with a {\it Fleiss'} $\kappa$ of $0.79$.
Taking the majority vote as ground truth, a total of $7,623$ articles were marked as clickbaits. The notable examples of articles  the volunteers marked non-clickbaits include the articles in the `news' section on Buzzfeed, most of which are reported like traditional news.

Finally, to have an equal representation of clickbait and non-clickbait articles while comparing them and building the classifier, we randomly selected $7,500$ articles from both the categories.

\vspace{-0.2cm}
\section{Comparing Clickbaits and Non-Clickbaits}
\vspace{-0.3cm}
\noindent We carried out a detailed linguistic analysis on the $15,000$ headlines both in the clickbait
and non-clickbait categories, using the Stanford CoreNLP tool~\cite{manning2014stanford}. 
A closer inspection of the clickbait headlines gives some insight about 
the semantic and syntactic nuances that occur more frequently in clickbait headlines 
compared to the traditional non-clickbait headlines.

\begin{figure}[t]
  \includegraphics[width=0.9\columnwidth,height=4.5cm]{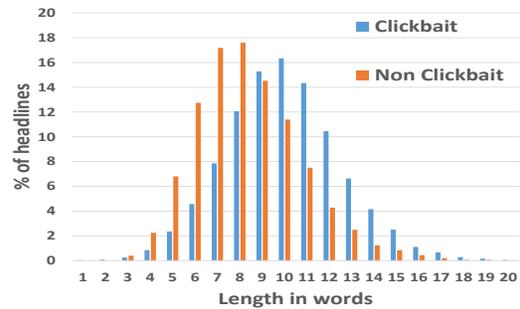}
  \vspace{-0.1cm}
  \caption{{Distribution of the length of both clickbait and non-clickbait headlines}}
  \label{fig:lengths}
  \vspace{-0.4cm}
\end{figure}

\vspace{-0.2cm}
\subsection{Sentence Structure}
\vspace{-0.3cm}
\noindent \textbf{Length of the headlines: } Figure~\ref{fig:lengths} shows the distribution
 of the number of words in both clickbait and non-clickbait headlines. It indicates that the conventional 
 non-clickbait headlines are shorter than clickbait headlines. 
 For example, the average length of the clickbait headlines 
 is $10$, whereas the average length is $7$ for non-clickbait headlines.
  
Traditional news headlines typically contain mostly \textit{content words} referring 
to specific persons and locations,
while the \textit{function words} are left out for readers to interpret from context. As an example,
consider the news headline: \textit{``Visa deal or no migrant deal, Turkey warns EU''}.
Here most of the words are content words summarizing the main takeaway from the story,
and it has very few connecting function words in between the content words. 

On the other hand, clickbait headlines are longer, well-formed English sentences 
that include both content and function words. One example of
such headlines is \textit{``A 22-Year-Old Whose Husband And Baby Were Killed By A 
Drunk Driver Has Posted A Gut-Wrenching Facebook Plea''}. It contains a generous mix of content and
function words.

\vspace{0.1cm}
\noindent \textbf{Length of the words: } Even though the number of words are more in 
clickbait headlines, the average word length is shorter. Specifically in our dataset, 
the average word length of clickbait headlines is found to be $4.5$ characters, while the
average word length of non-clickbait headlines is $6$.

\begin{figure}[t]
  \includegraphics[width=0.9\columnwidth,height=4.5cm]{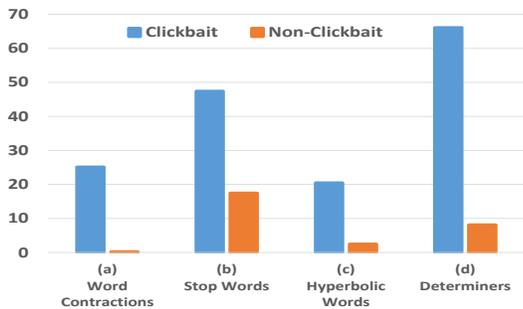}
  \caption{{Percentage of clickbait and non-clickbait headlines, which include (a) Word Contractions, (c) Hyperbolic Words, and (d) Determiners. (b) Percentage of words identified as stop words in both clickbait and non-click headlines.}}
  \label{fig:differences}
  \vspace{-0.5cm}
\end{figure}

The reason for shorter word lengths in clickbaits is primarily due to the frequent use 
of shorter function words and word shortenings. Shortened forms of words like \textit{\textbf{they're, you're, you'll, we'd}} 
are prevalent in clickbait headlines. 
On the other hand, they are not commonly found in non-clickbait headlines.
As we can see in Figure~\ref{fig:differences}(a), only $0.6\%$ of the traditional news 
headlines contain word shortenings, whereas nearly $22\%$ of clickbait headlines have such shortened words.
 
\vspace{0.1cm}
\noindent \textbf{Length of the syntactic dependencies: }
\if{0}
In the seminal work~\cite{tesniere1959elements}, Lucien Tesni{{\`e}}re argued that between any word and its neighbors in a 
sentence, the mind perceives \textit{connections} which form the structure of the sentence. 
Such structural connections establish \textit{dependency} relation between \textit{governor} and \textit{dependent} words. 
For example, in the sentence ``Birds fly'', the governor word is `fly' and the dependent word is `Birds'.
\fi
We used the Stanford collapsed-coprocessor dependency parser~\cite{de2006generating} 
to identify the syntactic 
dependencies between all pairs of words in the headlines, and then computed 
the distance between the {\it governing} and the {\it dependent words} in terms of the number 
of words separating them. 

\begin{figure}[t]
  \includegraphics[width=0.9\columnwidth,height=4.5cm]{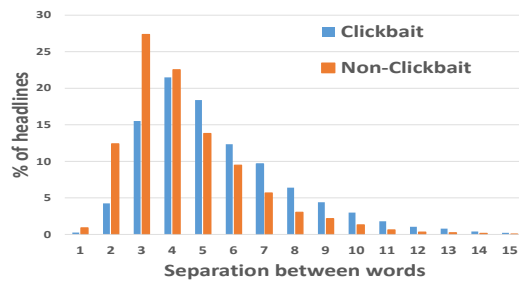}
  \caption{{Distribution of longest syntactic dependencies between all pair of words in clickbait and non-clickbait headlines.}}
  \label{fig:dependencies}
  \vspace{-0.5cm}
\end{figure}

Figure~\ref{fig:dependencies} shows the distribution of the maximum distance between 
governing and dependent words in both clickbait and non-clickbait headlines.
On average, clickbaits have longer dependencies than 
 non-clickbaits; the main reason being the existence of more complex phrasal sentences 
 as compared to non-clickbait headlines. Consider the example \textit{``A 22-Year-Old 
 Whose Husband And Baby Were Killed By A Drunk Driver Has Posted A Gut-Wrenching 
 Facebook Plea''}, where the subject `22-Year-Old' and the verb `Posted' are separated by 
 an adjective clause, making the length of the syntactic dependency as high as $11$.

\vspace{-0.2cm}
\subsection{Stop Words, Hyperbolic and Common Phrases}
\vspace{-0.3cm}
\noindent\textbf{Stop words: } Stop words are defined as the most common words that 
occur in any corpus of a particular
language. Figure~\ref{fig:differences}(b) shows the percentage of words present
 in both categories of headlines, which are stop words in English. It can be seen that, in clickbait
headlines, stop words are used more frequently (e.g. $45\%$ compared to $18\%$ in 
non-clickbaits) to complete the structure of the headlines.
On the other hand, in conventional news reporting, more \textit{content} words are used and 
inference of stop words is left to the reader. Due to this anomalous proportion of stop 
words in clickbait headlines and their contribution to sentence semantics, in the 
subsequent n-gram analysis, stop words were retained. 
This is a diversion from typical n-gram analysis, where stop words are removed before 
any analysis is performed.

\vspace{0.1cm}
\noindent\textbf{Hyperbolic words: } To compare the sentiment values of the constituent
words in both clickbait and non-clickbait headlines, we performed sentiment analysis using
the Stanford Sentiment Analysis tool~\cite{socher2013recursive}. We found that a
substantial fraction of clickbait headlines consist of words having `Very Positive' 
sentiments (e.g., \textit{Awe-inspiring, breathtakingly, gut-wrenching, 
soul-stirring, etc.}), which are almost non-existent in non-clickbait headlines.
 We call these extremely positive words as \textit{hyperbolic words}.
 Figure~\ref{fig:differences}(c) shows the percentage of headlines
 in both categories which include hyperbolic words. Use of such eye-catching words in clickbaits
strongly urge the reader to pursue the article with a promise of sensational information.

\vspace{0.1cm}
\noindent\textbf{Internet slangs: } Another class of words commonly found in clickbait
 headlines are Internet slang words like {\textit{WOW, LOL, LMAO, AMA, AF, etc}}.
Along with hyperbolic words, use of the slang words also immediately catches the
attention of the reader and lure them to read the article.

\vspace{0.1cm}
\noindent\textbf{Punctuation patterns: } Clickbait headlines also make use of informal
punctuation patterns such as {\textit{!?, ..., ***, !!!}} -- which are not used in 
conventional non-clickbait headlines.

\vspace{0.1cm}
\noindent\textbf{Common bait phrases: } Further, several commonly used catch phrases in
clickbait headlines exploit the ``curiosity gap" of users, such as \textit{``Will 
Blow Your Mind", ``You Won't Believe"}. We manually compiled a list of most commonly used bait 
phrases in the clickbait corpus. We further extended this list with the phrases used by
 Downworthy~\cite{downworthy} to detect clickbaits.

\vspace{-0.2cm}
\subsection{Subjects, Determiners and Possessives}
\vspace{-0.3cm}

\begin{table}
\begin{center}
\small
\begin{tabular}{|p{0.12\columnwidth}|p{0.77\columnwidth}|}
\hline
Clickbait & I, you, dog, everyone, girls, guys, he, here, it, kids, men, mom, one, parent, photos, reasons, she, something, that, they \\
\hline
Non-clickbait &  bomb, court, crash, earthquake, explosion, fire, government, group, house, U.S., China, India, Iran, Israel, Korea, leader, Obama, police, president, senate \\
\hline
\end{tabular}
\caption{{$20$ most commonly occurring subject words in both clickbait and non-clickbait headlines.}}
\label{table:commonSubjects}
\vspace{-0.5cm}
\end{center}
\end{table}

\noindent\textbf{Sentence subjects: } 
To identify the subject words in the headlines, we used the Stanford syntactic
dependency parser~\cite{de2006generating}, and then looked for the dependency relation
\textit{`nsubj'} among all the dependency relations found by the parser.
For example, the $20$ most commonly occurring subject words found in both clickbait and
non-clickbait headlines are listed in Table~\ref{table:commonSubjects}.

One interesting pattern we observed in clickbaits is the repetition of the popular subjects across 
many headlines. Nearly $62\%$ of the clickbait headlines contained one of the $40$ most common clickbait
 subject words. 
On the other hand, only $16\%$ of the non-clickbait headlines contained the top $40$ non-clickbait subject words.

\vspace{0.1cm}
\noindent\textbf{Determiners: } Clickbait headlines often employ determiners such as
 \textit{their, my, which, these} to reference particular people or things in the article.
Figure~\ref{fig:differences}(d) shows the percentage of headlines in both 
clickbait and non-clickbait headlines, where determiners are present. It can be seen that the use of determiners is 
way more in clickbaits compared to non-clickbaits.
The use of such determiners is primarily to make the user inquisitive about the object 
being referenced and persuade them to pursue the article further.

\vspace{0.1cm}
\noindent\textbf{Possessive case: } Clickbait headlines often address the reader in the 
first and second person with the use of subject words \textit{I, We, You}. 
Even the third person references are common nouns like \textit{he, she, it, they, man, dog} 
rather than specific proper nouns. This is in stark contrast with the non-clickbait headlines,
where the reporting is always done in third person.

\vspace{-0.2cm}
\subsection{Word N-grams, Part of Speech Tags and \\ Syntactic N-grams}
\vspace{-0.3cm}

\noindent\textbf{Word N-grams: } Word N-gram is defined as a contiguous sequence of N words 
from a given text sample, where N can vary from $1,2,3,...,$ to the length of the sample. 
From the clickbait and non-clickbait headlines, we extracted all possible $1$, $2$, $3$, and $4$-grams. 
For example, some commonly occurring $3$ and $4$-grams in clickbait headlines are
{\textit{`how well do you', `can we guess your', `what happens when', `how many 
of these'}}; while in non-clickbait headlines, some commonly occurring $3$ and $4$-grams are
{\textit{`dies at age', `kills at least', `us supreme court', `found guilty of', `won the match'}}.
We have found that the proportion of headlines that contained the top $0.05\%$ of unique n-grams in clickbaits was $65\%$, 
whereas in non-clickbaits, the proportion was $19\%$. 
Clickbait headlines follow a pattern of phrases repeatedly, while news headlines are inherently factual and 
unique in their reporting.

\vspace{0.1cm}
\noindent\textbf{Part of Speech Tags: } We tagged the headlines in both categories with 
the Part of Speech (POS) tags of their constituent words using the $45$ Penn Treebank POS
 tags~\cite{santorini1990part}. We specifically used the MaxEnt POS
 Tagger~\cite{ratnaparkhi1996maximum}. As most of the article headlines use 
title casing, we found some words to be erroneously tagged by the direct application of the POS tagger.

\begin{figure}[t]
  \includegraphics[width=0.98\columnwidth,height=5cm]{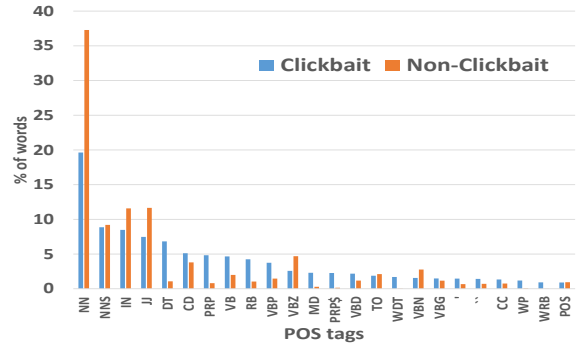}
  \caption{{Distribution of Parts-of-Speech (POS) tags for words in both clickbait and non-clickbait headlines}}
  \label{fig:pos}
  \vspace{-0.5cm}
\end{figure}

To overcome this limitation, we added a preprocessing step, where
we identified named entities using the Stanford Named Entity Recognizer
 tool~\cite{finkel2005incorporating} and retained those words in title case.
We converted every other word to lowercase to avoid ambiguity for the POS tagger.
After the preprocessing step, the POS tagger identified the POS tags for all words 
in the headlines. 

Figure~\ref{fig:pos} shows the distribution of POS tags.
From Figure~\ref{fig:pos}, we make the following interesting observations:

(i) Conventional non-clickbait headlines contain much larger proportion of proper nouns 
(POS tag: NN), indicating more content words and entities, than in clickbaits.

(ii) Clickbait headlines contain more adverbs and determiners (POS tags: RB, DT, WDT) than non-clickbait headlines.
 
(iii) Clickbaits also have higher proportion of personal and possessive pronouns (POS tags: PRP
 and PRP\$) like \textit{her, his, its, you} compared to non-clickbaits.

(iv) Clickbaits and non-clickbaits use verbs in different ways. Overall number of verbs is more in clickbaits as they 
focus on forming well-formed sentences. Regarding the person and 
the tense of the verbs, non-clickbait headlines tend to use more verbs in past participle 
and 3rd person singular form (POS tags: VBN and VBZ), whereas clickbaits use mostly past tense and non-3rd person singular forms (POS tags: VBD and VBP).

\vspace{0.1cm}
\noindent\textbf{Syntactic N-grams: } Syntactic N-grams (SN-grams) are formed by traversing paths of length N along the syntactic tree obtained from the collapsed-coprocessor dependency parsing~\cite{de2006generating} of the headlines. 
We do a depth-first traversal of the syntactic tree, using every node as a source.
For example, in the sentence ``A 22-Year-Old Whose Husband And Baby Were Killed By A Drunk Driver Has Posted A Gut-Wrenching Facebook Plea", we extract the SN-gram ``nsubj, acl:relcl, nmod:agent" corresponding to ``22-Year-Old Posted Driver Killed".
This way, SN-grams combine the syntactic dependencies among the non-neighbor words in a headline.

The advantages of using SN-grams are two-fold: (i) SN-grams are fewer in number than 
word n-grams, and (ii) SN-grams help capture linguistic phenomena by linking words 
that may not be neighbors in the surface structure but are syntactically related. 
For instance, in the headlines ``Which Disney Song Are You Based On Your Zodiac Sign" and ``Which Badass Witch Are You Based On Your Birth Month", the syntactic bigram \textit{(dobj, det)} captures the pattern {\textit{``Which **** Based"}}. 
An estimate of the number of such SN-grams is the height of the syntactic parse tree,
which on an average was found to be $10.03$ for clickbait and $6.45$ for non-clickbaits in 
our dataset.

\vspace{-0.2cm}
\section{Classifying Headlines as Clickbaits}
\vspace{-0.2cm}
\noindent
The comparative analysis, described in the earlier section, indicates prominent linguistic and structural differences between the clickbait and non-clickbait headlines. 
We attempt to use these differences as features to classify article headlines into clickbait and non-clickbait categories.

\vspace{-0.3cm}
\subsection{Feature Selection}
\vspace{-0.2cm}
\noindent\textbf{Sentence Structure: }
To capture the structural properties of the headline as classifier features, we used the length of the headline, the average length of words, the ratio of the number of stop words to the number of content words and the longest separation between the syntactically dependent words of a headline.

\vspace{0.1cm}
\noindent\textbf{Word Patterns: }
Word level structural features that we included were the presence of cardinal numbers in the beginning of the headline, 
presence of unusual punctuation patterns and the number of contracted word forms employed in the headline.

\vspace{0.1cm}
\noindent\textbf{Clickbait Language: }
Features that capture the nuances of the language employed, especially in the clickbait headlines, include the presence of hyperbolic words, common clickbait phrases, internet slangs and determiners. To model the popularity of the subject word in clickbait headlines as a feature, we used the score of a multinomial Naive Bayes classifier over sentence subjects. The score represents the ratio of the probability of assigning clickbait class label to the probability of assigning non-clickbait 
class label, given the subject word of the sentence. Model parameters for the Naive Bayed classifier were estimated using both datasets.

\vspace{0.1cm}
\noindent\textbf{N-gram Features: }
Word N-grams, POS N-grams and Syntactic N-grams were used as features. N-gram feature space grows linearly with the size of the dataset.
In order to limit the number of N-gram features used based on their frequency of occurrence, we pruned the feature space efficiently by using the sub-sequence property 
and an APRIORI-like algorithm~\cite{furnkranz1998study}. Similar to the case with  subject words, we built three multinomial Naive Bayes classifiers for the three sets 
of pruned features, i.e, the Word N-grams, POS N-grams and Syntactic N-grams. The scores of these three auxiliary Naive Bayes classifiers were used as inputs (i.e., as features)
 to the main classifier.
 
 \begin{table*}
\begin{center}
\begin{tabular}{ |*{16}{c|} }
\hline
 & \multicolumn{5}{c}{SVM} & \multicolumn{5}{|c|}{Decision Tree}  & \multicolumn{5}{c|}{Random Forest}\\
\hline
Features Used & Acc. & Prec. & Rec. & F1 & roc-auc & Acc. & Prec. & Rec. & F1 & roc-auc & Acc. & Prec. & Rec. & F1 & roc-auc\\
\hline
Sentence Structure & 0.77 & 0.79 & 0.75 & 0.77 & 0.84 & 0.74 & 0.77 & 0.69 & 0.73 & 0.8 & 0.75 & 0.76 & 0.73 & 0.74 & 0.82\\
Word Patterns & 0.84 & 0.84 & 0.83 & 0.84 & 0.91 & 0.84 & 0.84 & 0.84 & 0.84 & 0.91 & 0.84 & 0.84 & 0.84 & 0.84 & 0.91\\
Clickbait Language & 0.86 & 0.88 & 0.82 & 0.85 & 0.88 & 0.86 & 0.88 & 0.82 & 0.85 & 0.90 & 0.86 & 0.88 & 0.82 & 0.85 & 0.90\\
N-gram Features & 0.89 & 0.92 & 0.85 & 0.89 & 0.9 & 0.89 & 0.92 & 0.85 & 0.89 & 0.91 & 0.89 & 0.92 & 0.85 & 0.89 & 0.91\\
\hline
All Features & {\bf 0.93} & {\bf 0.95} & {\bf 0.90} & {\bf 0.93} & {\bf 0.97} & 0.90 & 0.91 & 0.89 & 0.90 & 0.90 & 0.92 & 0.94 & 0.91 & 0.92 & 0.97\\
\hline
\end{tabular}
\caption{Performance of the clickbait classifier using different prediction models.}
\label{tab:classificationResults}
\vspace{-0.4cm}
\end{center}
\end{table*}

\vspace{-0.2cm}
\subsection{Classification Performance}
\vspace{-0.3cm}
\noindent
We used a set of $14$ features spanning the structural, word-level, N-gram and linguistic categories, as described in the previous section. 
We experimented with three prediction models: Support Vector Machines (SVM) with Radial Basis Function (RBF) kernel, Decision Trees, and Random Forests.
Table~\ref{tab:classificationResults}
shows the $10$-fold cross validation performance 
(specifically Accuracy, Precision, Recall, F1, and ROC AUC scores) for all three prediction models.
Table 2 details the evaluation scores if each category of features were used independently and also for the combined feature set. 
We can see from Table~\ref{tab:classificationResults} 
that SVM performed best with an accuracy of $93\%$, a precision of $0.95$ and recall of $0.9$. 

Finally, as a baseline for comparison, we took the fixed set of rules employed by Downworthy~\cite{downworthy} to detect clickbaits, 
and ran it on our dataset. The $10$-fold cross validation performance achieved by Downworthy is $76\%$ accuracy
 with $0.72$ recall and $0.78$ precision. 
Therefore, the proposed classification technique outperforms the baseline with a large margin.

Further note that, even though the current classifier works with English headlines only, the characteristics of clickbait used 
as features here are common linguistic phenomena occurring across languages and hence,
the classifier can easily be extended to other languages.

\vspace{-0.2cm}
\section{Blocking Clickbait Headlines}
\vspace{-0.3cm}
\noindent In the previous section, we observed that the classifier achieves $93\%$ accuracy in detecting
 clickbaits. With this impressive performance
of the classifier, the next logical step is to devise an approach which can block certain clickbaits 
according to the reader's discretion.

Towards that end, we attempted to understand whether there is some uniformity in choice about
the type of headlines readers would like to block.
We conducted a survey by asking a group of $12$ regular news readers to review $200$ randomly chosen clickbait headlines.
The readers were asked to mark the headlines they would have clicked while surfing the web, and also mark the headlines they 
would have liked to block.
We then computed the average Jaccard coefficient between all pairs of readers, for the headlines to be 
clicked as well as the headlines to be blocked.
Though all $200$ headlines show the characteristic features of clickbaits, the average Jaccard coefficients for
clicked as well as blocked headlines are $0.13$ and $0.15$ respectively. These low Jaccard scores indicate
that different readers interact with different
clickbait headlines very differently, and there is not much commonality in reader behaviors towards clickbait headlines.
Effectively, readers' interpretation of clickbait is subject to their own interests, likes as well as dislikes.

\vspace{0.1cm}
\noindent\textbf{Blocking as personalized classification: }
As readers' interaction with clickbait headlines vary drastically, a one-size-fits-all approach can not work in blocking 
clickbaits. We instead need personalized classifiers for individual readers to classify the clickbaits into the ones to 
block and the ones not to block.
Essentially, the problem translates to modeling the reader's interests from the articles she has read 
as well as blocked in the past. 
Accordingly, for a new clickbait, we need to predict whether the reader would like to
block this new article or not.

The notion of reader interests in clickbait articles or lack thereof can be modeled with the help of two interpretations. 
For instance, if some reader decides to block the clickbait headline \textit{``Can You Guess The Hogwarts House Of These 
Harry Potter Characters?''}, we can make two conclusions. 
(i) The reader may not be interested in the topic `Harry Potter' itself, and does not want to read any article related to `Harry Potter';
or (ii) she may be annoyed by the commonly occurring pattern \textit{``Can You Guess ..."} but may click on another 
`Harry Potter' related article in the future. There can also be cases, where both reasons play a role. 
Hence, we modeled and designed methods to capture both notions of reader interests as well as a combination of both factors.

\vspace{0.1cm}
\noindent\textbf{Blocking based on topical similarity: }
Our first approach to block clickbaits is to first extract a set of topics from an article with clickbait headline, 
and find the similarity between this set and the topics previously extracted from blocked and clicked articles.

To find the topics of interest to a reader, we chose to use the content words in the article headline, 
as well as the article metatags and keywords that occur in the $<$head$>$ part 
of articles' html sources.
For instance, in the html source for the article having headline
\textit{``We Tried The New Starbucks Butterbeer Drink And Dumbledore Would Definitely Approve"}, 
the tags found were \textit{butterbeer, harry potter, hermione, jk rowling, wizarding world}. 
These tags are given by the developer of the corresponding webpages and they contain topical information regarding
the article.
They naturally identify the hidden topics of an article that other topic-modeling systems like 
LDA
\footnote{LDA here performs poorly as the available data are sparse and noisy.}  
will take large training data to identify.
Tags and keywords extracted from a particular clickbait link are stored as attributes \textit{ClickTags} or \textit{BlockTags},
 depending on whether the link has been clicked or blocked.

For a given link, we used BabelNet~\cite{navigli2012babelnet}\footnote{BabelNet~\cite{navigli2012babelnet} is a multilingual semantic network which connects $14$ million concepts and named entities extracted from WordNet and Wikipedia. Each node in the network is called a BabelSynset.} to expand its \textit{BlockTags} or \textit{ClickTags} sets. We discover the nodes in BabelNet that correspond to these tags. These nodes initially form a self-contained cluster, called a \textit{Nugget}. We expand the nugget further by adding common hypernyms of member nodes. Two Nuggets are merged when a BabelSynset (i.e. a node) is common to both. 

Considering all the article links a reader has blocked and clicked respectively, we form a reader's \textit{BlockNuggets} and \textit{ClickNuggets}. Then for a new clickbait link, the block/do not block decision is predicted based on whether the
nugget for the new link is more \textit{similar} to \textit{BlockNuggets} or \textit{ClickNuggets}. 
Here similarity is computed
based on the number of nodes common in two nuggets.

For every reader, we use the top`$100$ blocked and clicked links ordered by timestamps for tag extraction and nugget formation. This is done to limit the data considered for training to the latest reader interests, which can change with time. 

\vspace{0.1cm}
\noindent\textbf{Blocking based on linguistic patterns: }
In the second approach, we identified the linguistic patterns in the articles that the reader
clicks on or chooses to block. The pattern is formed by normalizing the words in the 
headlines in the following ways. (i) Numbers and Quotes are replaced by tags 
$<$D$>$ and $<$QUOTE$>$. (ii) The top $200$ most commonly occurring words in the clickbait corpus, 
including English stop words, were retained in their original form. 
(iii) Content words such as Nouns, Adjectives, Adverbs and Verb inflections were replaced 
by their POS tags. For instance, ``Which Dead `Grey's Anatomy' Character Are You", reduces to ``which JJ $<$QUOTE$>$
 character are you" and 
``Which `Inside Amy Schumer' Character Are You" reduces to ``which $<$QUOTE$>$
 character are you". 
 
We convert each headline into such patterns, and thus we get a set of patterns for
both blocked articles and clicked articles. To compute the similarity between two patterns, 
we use the {\it word-level edit distance}. 
Using the mechanism similar to the topical similarity case, we make the block/do not block decision. 

\vspace{0.1cm}
\noindent\textbf{Hybrid approach: } We also experimented with a hybrid approach which takes into account both 
topical similarity and linguistic patterns. For a new article, its tags are extracted, nugget is formed and compared with
the \textit{BlockNuggets} and \textit{ClickNuggets} -- this gives the topical similarity scores. Similarly, we get the linguistic
similarity scores. The hybrid scores are obtained using a weighted combination of both topical and linguistic similarity scores, 
and finally we make the block/do not block decision based on the hybrid scores.

\vspace{0.1cm}
\noindent \textbf{Evaluation: } We tested all three approaches using the click and block decisions marked by the $12$ readers.
 Table~\ref{tab:personalResults} shows the average accuracy, precision, recall and F1 scores for all approaches. 
Note that here in the hybrid approach, we have taken equal weights of $0.5$ for both topical and linguistic similarity scores. 
Exploring the effects of other weights is left as future work.
 
From Table~\ref{tab:personalResults}, it is evident that the pattern
 based approach yields better results. It also executes faster compared to the more involved topic based approach, therefore, it is better suited for a
 real-time environment. Hence, it was integrated into the browser extension `Stop Clickbait' which we will discuss in the next section.

\begin{table}
\begin{center}
\small
\begin{tabular}{ |*{5}{c|} }
\hline
Approach & Accuracy & Precision & Recall & F1 \\
\hline
Pattern Based & {\bf 0.81} & {\bf 0.834} & {\bf 0.76} & {\bf 0.79} \\
\hline
Topic Based & 0.75 & 0.769 & 0.72 & 0.74 \\
\hline
Hybrid & 0.72 & 0.766 & 0.682 & 0.72 \\
\hline
\end{tabular}
\caption{{Performance of different blocking approaches.}}
\label{tab:personalResults}
\end{center}
\vspace{-0.5cm}
\end{table}

\vspace{-0.2cm}
\section{Browser Extension: Stop Clickbait}
\vspace{-0.3cm}
\noindent In the earlier sections, we showed that both the classifier and the blocking approach achieve
high accuracy in detecting and blocking clickbaits. Hence, to increase the applicability of the proposed approaches,
and to help the users to deal with clickbaits across different websites,  
we attempt to build an extension `Stop Clickbait' for Chrome browsers.
`Stop Clickbait' warns the users about the existence of clickbaits in 
different webpages. It provides the users the facility to block certain clickbaits whereby it
automatically block similar clickbaits in future visits. We next describe the working of
the chrome extension.

When a webpage is loaded in Chrome, `Stop Clickbait' scans the {\it Document 
Object Model (DOM)} for anchor elements ($<$a href = ... $>$), and it also keeps listening 
for dynamic insertion of anchor elements in the DOM. 
Once an anchor tag is discovered, it proceeds to check if the anchor tag has any anchor text.
If the anchor text is available, it is sent to a server, 
 where the text is input to the clickbait classifier.
If the anchor text is not available in the DOM, then the url is sent to the server, where it makes a GET request for the 
webpage title and runs the classifier on the obtained title.

\begin{figure*}[tb]
\center{
\subfloat[{\bf }]{\includegraphics[width=0.32\textwidth,height=4.5cm]{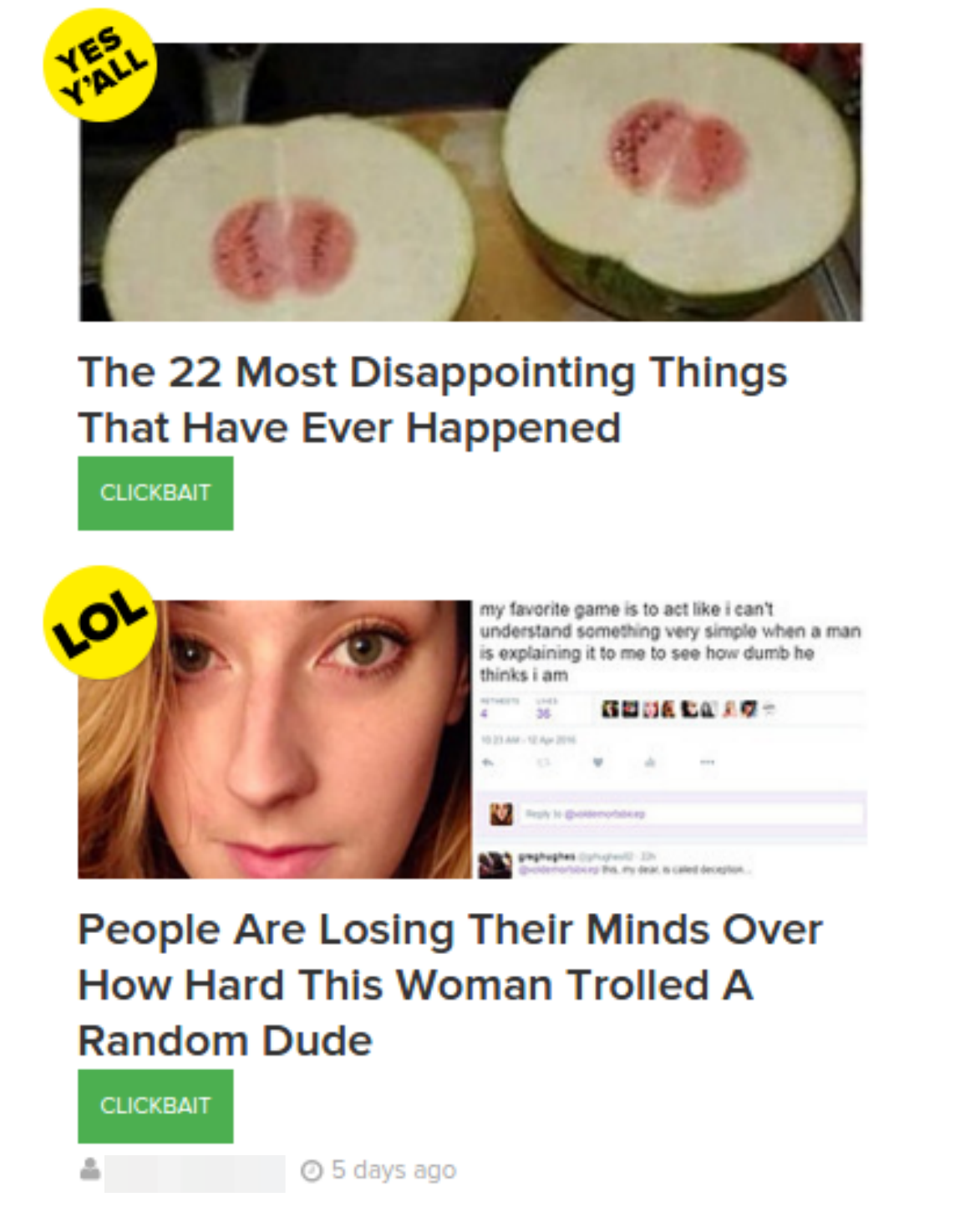}}
\hfil
\subfloat[{\bf }]{\includegraphics[width=0.32\textwidth,height=4.5cm]{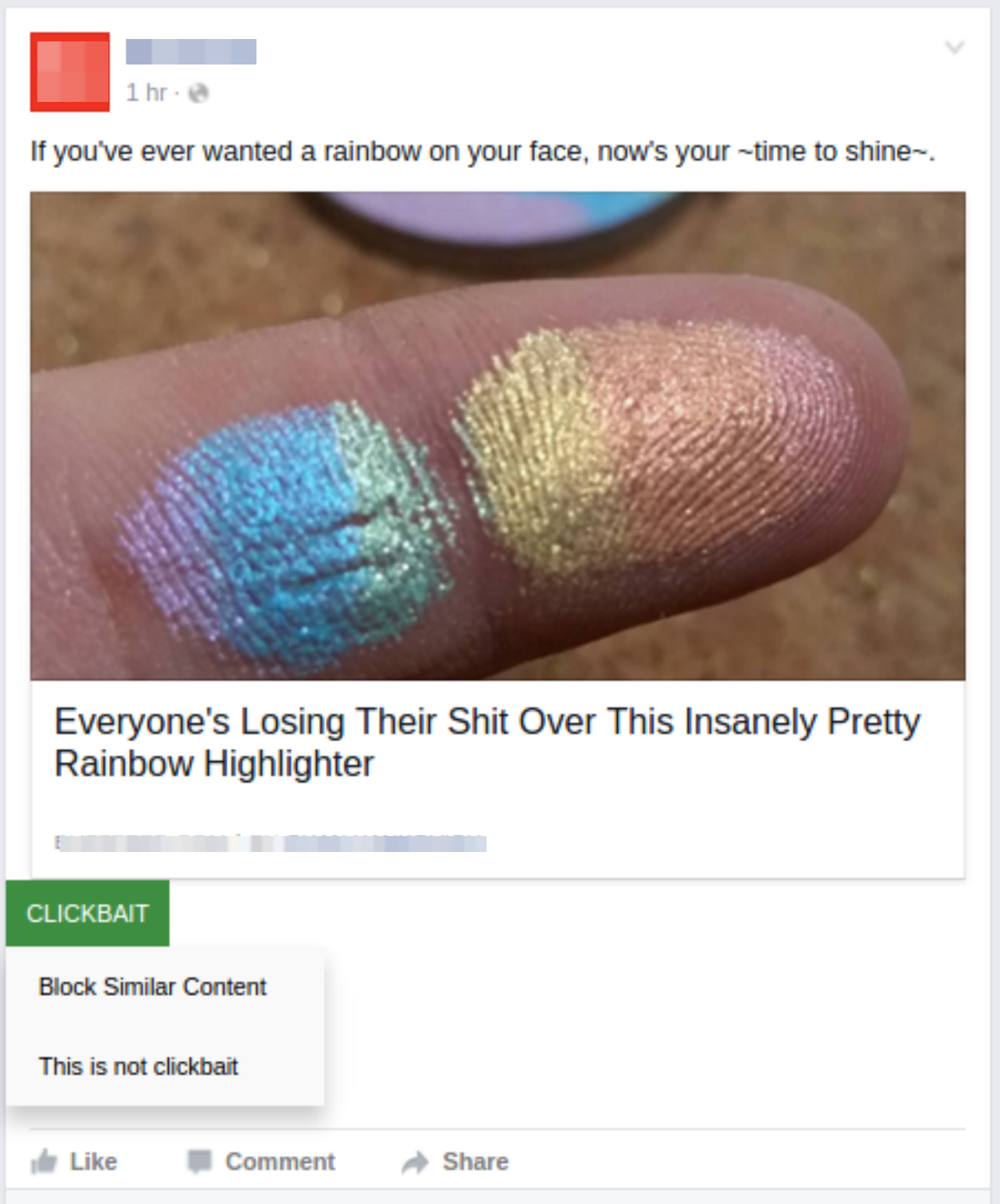}}
\hfil
\subfloat[{\bf }]{\includegraphics[width=0.32\textwidth,height=4.5cm]{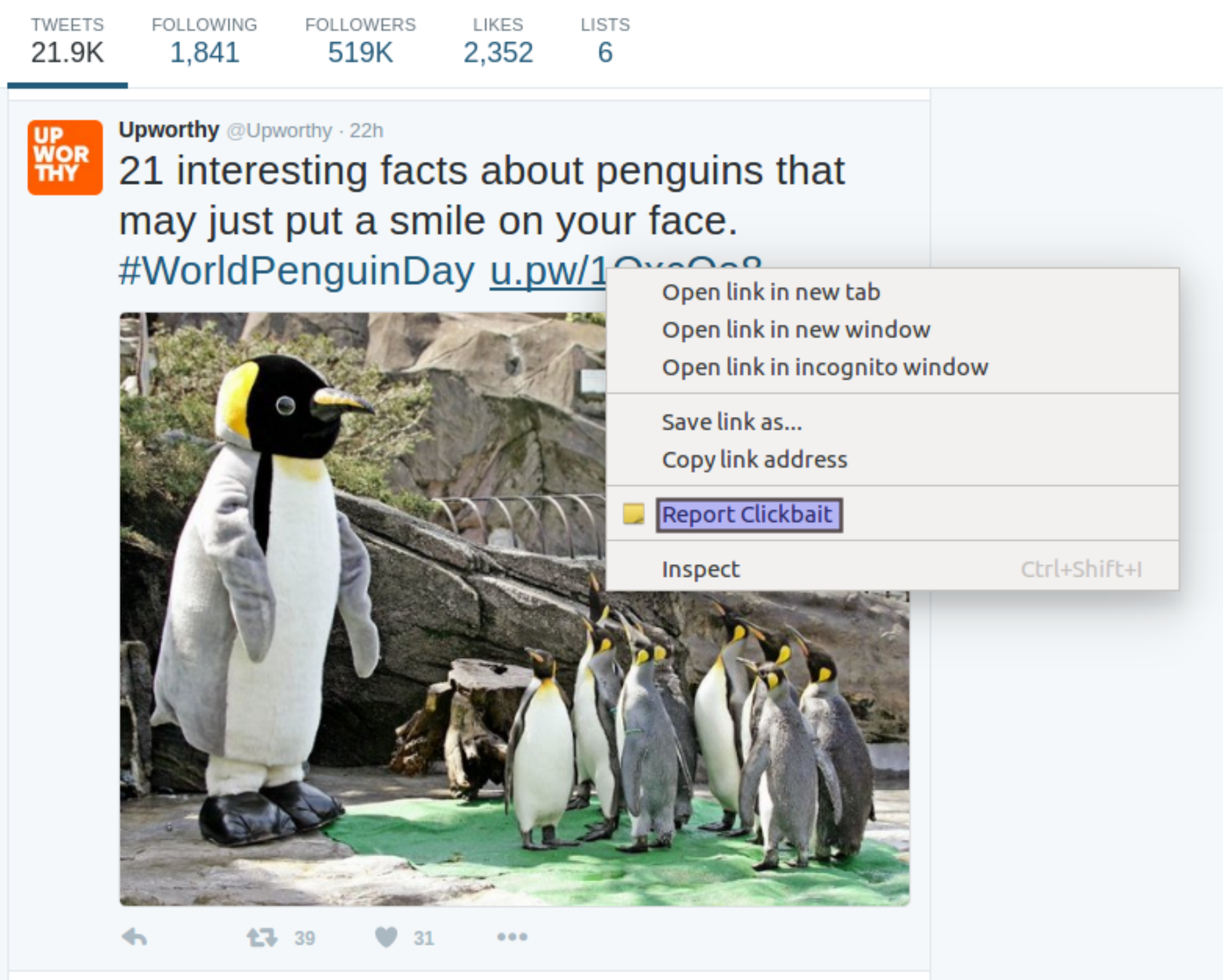}}
}
\caption{{Different snapshots of the working of `Stop Clickbait' extension: (a) Green clickbait indicators in webpages, (b) Option to block a link or report misclassification, (c) Option to report link which should be marked as clikbait.}}
\label{fig:extension}
\vspace{-0.4cm}
\end{figure*}

Then, based on the anchor text or the webpage title, the classifier predicts whether the link
is clickbait or not. The result of the classification is fed back to the extension, and 
links that are flagged as clickbaits is marked with a green button (by adding 
an element into the DOM) and links that are not clickbaits is left unmarked. 
Figure~\ref{fig:extension}(a) shows the green clickbait indicators in one of the webpages.
Finally, when the user clicks on the green indicator button (as shown in Figure~\ref{fig:extension}(b)), two options appear for the user: (i) block similar content in future, and 
(ii) the clickbait indicator was placed erroneously, i.e. the classification
was wrong. Further, in case, the extension fails to mark a genuine clickbaits in a webpage, the user can click on the link (as shown in Figure~\ref{fig:extension}(c)), and mark it as clickbait. We store these feedbacks in the server and the clickbait classifier is retrained every day with the added feedback data.

\vspace{0.1cm}
\noindent\textbf{Evaluating the performance of the extension: }
We uploaded the extension to the official Google chrome store\footnote{The extension is available at chrome.google.com/webstore/detail/stop-clickbait/iffolfpdcmehbghbamkgobjjdeejinma} and also circulated the extension binary in our peer groups. 
We provided a randomly generated unique 32 byte identification number for every instance of the extension to facilitate
training for each personalized classifier.

Overall $37$ people used the extension during the month of April, 2016. These users carried out their regular web 
activities and visited different websites such as Facebook, Buzzfeed, New York Times, and they reported both
false positives and false negatives. We present in Table~\ref{tab:extensionPerformance}, the overall performance of 
the extension across different websites and also individually at $5$ different domains. We can see that the classification performance is very good yielding $94\%$ accuracy and F1-score of $0.934$ across all websites the users visited.

Out of these $37$ users, $16$ users explicitly blocked more than $10$ clickbaits while browsing different websites. We 
invoked the personalized classifier as described in the earlier section and blocked different clickbaits during
 their further visits. We also provided an option to
 the users to check the links which have been blocked by the extension and give feedback
on whether the blocking was a right decision or not. According to the user feedbacks, on average, the extension had
 correctly blocked $89\%$ of the links.

\begin{table}
\begin{center}
\small
\begin{tabular}{ |*{5}{c|} }
\hline
Media Website & Accuracy & Precision & Recall & F1 \\
\hline
Huffington Post	& 0.92 & 0.88 & 0.93 & 0.904 \\
\hline
CNN	& 0.97 & 0.93 & 0.99 & 0.96 \\
\hline
Buzzfeed & 0.98 & 1.00 & 0.97 & 0.984 \\
\hline
New York Times & 0.95 & 0.83 & 0.95 & 0.88 \\
\hline
Facebook & 0.93 & 0.85 & 1.00 & 0.92 \\
\hline
Overall & {\bf 0.94} & {\bf 0.92} & {\bf 0.95} & {\bf 0.934} \\
\hline
\end{tabular}
\caption{Performance of the extension at different sites.}
\label{tab:extensionPerformance}
\vspace{-0.5cm}
\end{center}
\end{table}

\vspace{-0.3cm}
\section{Related Work}
\vspace{-0.3cm}
\noindent
The origin of clickbaits can be traced back to the advent of tabloid journalism, which started focusing on `soft news' compared
to `hard news', and sensationalization rather than reporting in depth and truthful account of the events. There has been 
many research works in media studies highlighting the problems with tabloidization. For example, Rowe~\cite{rowe2011obituary} 
examined how the common tabloid properties like simplification and spectacularization of news, are making its way into 
the more conventional newspapers and how it is changing the course of professional journalism. Similarly, the current concerns 
on the prevalence of clickbaits~\cite{clickbait_bad} highlight the changing face of journalistic gatekeeping during the  
abundance of clickbait articles having very low news value.

There has been recent works to understand the psychological appeal of clickbaits. Blom et. al.~\cite{blom2015click} examined how clickbaits employ 
two forms of forward referencing -- {\it discourse deixis} and {\it cataphora} -- to lure the readers to click on the article links.
Chen et. al.~\cite{chen2015misleading} argued for labeling clickbaits as misleading content or false news.

However, there has been little attempt to detect and prevent clickbaits. As mentioned earlier, Facebook attempted to remove clickbaits depending on the click-to-share ratio and the amount of time spent on different stories. A recent work by Potthast et al.~\cite{potthast2016clickbait} attempted to detect clickbaity Tweets in Twitter by using common words occurring in clickbaits,
and by extracting some other tweet specific features. The browser extension `Downworthy'~\cite{downworthy} detects clickbait 
headlines using a fixed set of common clickbaity phrases, and then converts them to meaningless garbage text. 
The problems with the above approaches are that they either work on a single domain, or the fixed ruleset does not capture the nuances employed across different websites.
In this work, we propose and demonstrate a comprehensive solution which works very well across the web.

\vspace{-0.2cm}
\section{Conclusion}
\vspace{-0.3cm}
\noindent
In this paper, we compared clickbait and non-clickbait headlines, and highlighted many interesting differences between these two categories. 
We then utilized these differences as features to detect clickbaits. We also proposed personalized approaches which can 
block certain clickbaits according to reader interests. Finally, using these two components, we have developed a Chrome extension which warns the readers of different media websites about the presence of clickbaits in these websites. The extension also gives
the option to the readers to block clickbaits and it automatically blocks similar ones in subsequent visits by the readers.

To the best of our knowledge, our work is the first attempt to provide a comprehensive solution to deter the prevalence of clickbaits. However, the job is far from over. Our future work lies in improving the classification and blocking performances further and tune the extension according to further user feedback.
Finally, it is our belief that combating the prevalence of clickbaits should be a community initiative and towards that end, 
we have made the data and source codes publicly available at \textbf{\url{cse.iitkgp.ac.in/~abhijnan}}, so that the researcher and the developer communities can come forward, and collectively make the effort a grand success.

\vspace{0.1cm}
\noindent\textbf{Acknowledgments:} 
A. Chakraborty was supported by Google India PhD Fellowship and the Prime Minister's Fellowship for Doctoral Research.

\vspace{-0.25cm}
\bibliographystyle{IEEEtran}
\small
{
\bibliography{Main}

\begin{thebibliography}{10}
\bibitem{clickbait_bad}
J.~Dvorkin, ``Column: Why click-bait will be the death of journalism,''
  pbs.org/newshour/making-sense/what-you-dont-know-about-click-bait-journalism-could-kill-you/.
\end{thebibliography}
}
\end{document}